\documentclass[sigconf]{acmart}

\usepackage{url}
\usepackage[justification=centering]{caption}

\AtBeginDocument{%
  \providecommand\BibTeX{{%
    \normalfont B\kern-0.5em{\scshape i\kern-0.25em b}\kern-0.8em\TeX}}}

\setcopyright{acmcopyright}
\copyrightyear{2022}
\acmYear{2022}
\acmDOI{10.1145/1122445.1122456}

\acmConference[ACM ITiCSE '22]{$27^{\text{th}}$ Annual Conference on Innovation and Technology in Computer Science Education (ITiCSE)}{July 11--13, 2022}{Dublin, Ireland}

\begin{document}

\title{Towards Understanding the Skill Gap in Cybersecurity}

\author{Francois Goupil}
\authornote{Both authors contributed equally to this research.}
\email{francois.goupil@uni.li}
\author{Pavel Laskov}
\authornotemark[1]
\email{pavel.laskov@uni.li}
\affiliation{%
  \institution{University of Liechtenstein}
  \streetaddress{Fürst Franz Josef Strasse}
  \city{Vaduz}
  \country{Liechtenstein}
  \postcode{9490}
}

\author{Irdin Pekaric}
\authornote{Both authors contributed equally to this research.}
\email{irdin.pekaric@uibk.ac.at}
\author{Michael Felderer}
\authornotemark[2]
\email{michael.felderer@uibk.ac.at}
\affiliation{%
  \institution{University of Innsbruck}
  \streetaddress{}
  \city{Innsbruck}
  \country{Austria}
  \postcode{6020}
}

\author{Alexander Dürr}
\authornote{Both authors contributed equally to this research.}
\email{alexander.duerr@uni-wuerzburg.de}
\author{Frederic Thiesse}
\authornotemark[3]
\email{frederic.thiesse@uni-wuerzburg.de}
\affiliation{%
  \institution{University of Würzburg}
  \streetaddress{}
  \city{Würzburg}
  \country{Germany}
  \postcode{}
}

\begin{abstract}
Given the ongoing "arms race" in cybersecurity, the shortage of skilled professionals in this field is one of the strongest in computer science. The currently unmet staffing demand in cybersecurity is estimated at over 3 million jobs worldwide. Furthermore, the qualifications of the existing workforce are largely believed to be insufficient. We attempt to gain deeper insights into the nature of the current skill gap in cybersecurity. To this end, we correlate data from job ads and academic curricula using two kinds of skill characterizations: manual definitions from established skill frameworks as well as "skill topics" automatically derived by text mining tools. Our analysis shows a strong agreement between these two analysis techniques and reveals a substantial undersupply in several crucial skill categories, e.g., software and application security, security management, requirements engineering, compliance and certification. Based on the results of our analysis, we provide recommendations for future curricula development in cybersecurity so as to decrease the identified skill gaps.  

\end{abstract}


\maketitle

\section{Introduction}

Cybersecurity is not only a ``gold mine'' for criminals but also one of the most lucrative fields for job seekers---if they qualify.

``Nowhere is the workforce-skills gap more pronounced than in cybersecurity", concluded the World Economic Forum in 2017 and projected 3.5 million unfilled jobs in cybersecurity by 2021 \cite{wef:2017}. This forecast turned out to be quite accurate, as the 2020 (ISC)\textsuperscript{2} Cybersecurity Workforce Report revealed a current gap of 3.1 million jobs \cite{isc2:2020}. Furthermore, the recent State of Cybersecurity Report of the Information Systems Audit and Control Association (ISACA) \cite{isaca:2021} claims that over 60\% of cybersecurity teams in the USA are understaffed and 50\% of cybersecurity job applicants are not well qualified. 

These facts expose two facets of the cybersecurity skill gap: (a) the sheer number of unfilled positions and (b) the discrepancy between the skills needed in the industry and the job seekers' qualifications. The latter is of particular importance for computer science education which shapes the future workforce. In this paper, we attempt to obtain a deeper understanding of the qualifications discrepancy in cybersecurity and propose techniques for a routine execution of such analyses. 

A traditional approach for the characterization of skills involves manual development of a ``skill framework'' reflecting knowledge, roles, position level, etc. Examples of skill frameworks related to cybersecurity are the ACM Computing Classification System (CCS, "Security and Privacy" category) \cite{ACM}, the Cybersecurity Body of Knowledge (CyBOK) \cite{cybok} and the EU Cybersecurity Taxonomy \cite{JRC}. A skill framework provides comprehensive coverage of various educational aspects. Its development, however, requires substantial manual work and cannot be easily updated through time.

In the recent decade, various text mining tools have been proposed for \emph{automatic skill characterization} in different industries, e.g., \cite{aken2009mining,debortoli2014comparing,fall2018identifying}, as well as for exploring academic curricula \cite{foll2021exploring}. Text mining enables construction of topic models from the underlying data. Such models represent topics as sets of keywords (which can be manually annotated) and assess the importance of each topic in the overall data corpus. When text mining tools are applied to the job ad or curricula data, the resulting topics are likely to characterize the skills referred to by the respective documents. Such characterizations can then be used for the analysis of skill gaps. While topic models cannot capture semantic relationships between various qualification-related aspects as skill frameworks do, their obvious advantage is the low amount of manual work and hence, the possibility for a periodic re-assessment. 

In this contribution, we demonstrate that both manual and automatic analysis provide a consistent characterization of skill gaps in cybersecurity. Our investigation is based on the job ad and the academic curricula data collected in the German-speaking countries of Europe (Germany, Switzerland, Austria and Liechtenstein). The main instrument of our analysis is the alignment of document frequencies of various skill categories, i.e., the percentage of documents in the corpus containing a certain skill category. Skill categories with substantially different document frequencies in job ads and curricula represent a mismatch between an industrial demand and the academic supply. Such analysis enables us to identify skills of high demand in the industry but low supply in academic programs and vice versa. In the following, we refer to the shortage of certain skills in industry as a ``positive'' and the oversupply of skill by academia as a ``negative'' skill gap.  

The main difference between the manual and the automatic variants of our analysis lies in the construction of skill topics. Our manual analysis uses fuzzy matching to compute document frequencies for skill categories defined in existing skill frameworks. In our automatic analysis, the skill categories are extracted from the respective data as topics using text mining and are manually annotated as short skill descriptions. The document frequencies of topics for the automatic analysis are obtained as a by-product of the text mining step. 

The results obtained via our analysis offer interesting insights into the skill gaps in cybersecurity. We compare the skill categories for three experimental setups: automatic, manual using the ACM CCS, and manual using the EU Cybersecurity Taxonomy. Both positive and negative skill gaps can be reported based on our data corpus.  
The most significant positive skill gaps exist in software and application security and security management. On the other hand, certain classical topics of cybersecurity such as cryptography and cryptanalysis are not in high demand in industry any more, which raises a question whether they should still be considered as an important cornerstone of cybersecurity education. Finally, increasing industrial demand for various emerging cybersecurity skills such as cloud services, cyber threat intelligence and security compliance, as well as for certain important branches such as automotive systems and online banking has been pinpointed by the automatic analysis. The demand for such skill, albeit not too high in absolute terms, is so far hardly met by the academic programs. 

This paper is organized as follows. A brief account of related work is presented in Section~\ref{sec:related} as a motivation for our study. The data collection and processing methods are presented in Section~\ref{sec:methods}. The skill gaps identified as a result of our study are presented in Section~\ref{sec:gaps}. Implications of the identified skill gaps for computer security education are discussed in Section~\ref{sec:discussion}, followed by the summary and conclusions in Section~\ref{sec:conclusions}.

\section{Related Work}
\label{sec:related}

Computer science and information systems have always suffered from the lack of a qualified workforce. It is therefore not surprising that various prior works attempted to characterize the required skills and to analyze a discrepancy between industry expectations and educational outcomes. Early approaches to this problem, e.g., \cite{trauth1993expectation}, conducted surveys among managers and university professors aimed at devising a qualitative description of the IT skill portfolio in industry and academia.

A broad class of literature related to skill gap analysis addresses the task of building characterizations of essential skills in a certain field of computer science.  For example, \cite{aken2007impact} developed a catalog of 95 technical and non-technical skills believed to be essential for information systems professionals (in 2007) and surveyed HR managers to explicitly assess the skill gap in each category. Further examples of manual construction of skill definitions are the ACM CCS \cite{ACM}, CyBOK \cite{cybok} and the EU Cybersecurity Taxonomy \cite{JRC}.  

The rapid development of text mining and NLP techniques offers a powerful tool for automatic skill characterization. One of the earliest works of this kind matched a pre-defined set of related keywords against 244,460 unique job advertisements and used clustering to group most frequent skills into job profiles \cite{aken2009mining}. Text mining approaches have also been used to compare skill sets in different branches of the IT industry \cite{debortoli2014comparing}, mapping skills to different professional roles \cite{fall2018identifying}, design and analysis of curricula \cite{foll2021exploring}, as well as for longitudinal studies of industrial skill demands \cite{handali2020industry}, to name only a few exemplary works. An example of a hybrid approach combining surveys with job ad and curricula analysis based on a manual skill taxonomy can be found in recent work on skill requirements in digital forensics \cite{hranicky2021dfrws}.

In comparison with prior work our approach exhibits the following distinctive features:

\begin{itemize}
\item We present a quantitative assessment of the skill gaps as the discrepancy in document frequencies between similar skills in job ads and curricula.
\item We verify that manual and automatic skill extraction delivers coherent results, the latter being capable of detecting novel skill demands. 
\item We demonstrate that a multilingual nature of job ad and curricula data can be effectively handled by modern text mining tools.
\item Using this novel methodology, we analyze the skill gap in the field of cybersecurity and make recommendations for the future curricula development. 
\end{itemize}

\section{Methods}
\label{sec:methods}

In this section, we provide a detailed account of data collection and analysis methods. Specifically, we present the methods for the collection of the job ad and curriculum data as well as the techniques for manual and automatic extraction of skill categories. 

\subsection{Collection of Job Advertisement Data}

The job-ad data is the main source of information about the skills expected by the cybersecurity industry. 
 To build such a data corpus, we developed web crawlers for the scraping of job ads at Indeed, Stepstone, and Monster---three job portals extensively used in the German-speaking countries. Note that the data-driven skill gap analysis inevitably leads to a multilingual environment. Especially for the cybersecurity workforce, which is highly international, understanding of applicants' skills regardless of the potential employer's locations as well as of the language of her education is highly desirable. Hence, the design of our study provides a perfect setup for the investigation of multilingual aspects in skill analysis. By presenting the results obtained from the mixture of English and German data we ensure the transferability of our analysis procedure to other countries. 

Our web crawlers were built using the \textit{scrapy} library and entail the following  main steps:

\begin{enumerate}
    \item Querying the aforementioned three job portals for keywords “cyber-security” and “it-sicherheit” (cyber-security in German) and collecting the resulting URLs.
    
    \item Crawling each URL after a three-second timeout and retrieving the job id, title, location, company, description as well as country.
    
    \item Storing this data in a PostGreSQL database.
\end{enumerate}
 In order to obtain a representative dataset, we ran the scrapers every second Sunday for 3 months starting from 5 am until 1 pm.

Subsequently, we performed various data cleaning and filtering tasks to improve our dataset quality. We first removed duplicates within each single job portal (arising when the same job ad is encountered during a different time interval) using the respective job IDs. Then, using the hash of the job description, we also removed duplicates originating from different job portals, as it is not uncommon for companies to publish the same job ad on different portals. We found that looking for the cybersecurity keywords was not enough to guarantee that a job ad is indeed related to cybersecurity. Therefore, we filtered the data using a more restrictive matching rule: the keyword “security” (German: “Sicherheit”) should either occur in the job title or at least three times in the job description. 

Our data collection procedure yielded 933 job ads in English and 1,530 job ads in German.  As it can be seen in Table~ \ref{tab:job ads}, this dataset exhibits a good coverage of the respective regions and languages despite its relatively small size. 
\begin{table}[h!]
\centering
\setlength\tabcolsep{1.5pt}
\begin{tabular}{|l|l|l|l|l|l|l|}
\hline
Country (language) & S (ger) & S (en) & A (ger) & A (en) & G (ger) & G (en) \\ \hline
Number of job ads & 214 & 272 & 226 & 158 & 1028 & 503 \\ \hline
\end{tabular}
\caption{Distribution of job ads per country and language: Switzerland (S), Austria (A), Germany (G)}
\label{tab:job ads}
\vspace{-1.5\baselineskip}
\end{table}

\subsection{Collection of Curricula Data}

The second dataset needed for our approach contains the cybersecurity curricula in higher education institutions in the German-speaking countries. It covers all explicitly stated cybersecurity study programs at selected public universities and universities of applied sciences. To keep the dataset balanced across the three countries, a representative subset of German universities was chosen, whereas in Austria and Switzerland all universities were considered. Data acquisition comprised the following steps: (1) investigation of the relevant sources, (2) gathering of the raw data, and (3) extraction of curricula descriptions.

In the initial stage, it was necessary to investigate the building blocks of curricula in each country. These included the course information covering name, description, university, and level. The level shows if the course is included in bachelor or master studies. The search for relevant curricula was done manually by checking official lists and databases of universities and their study programs.  For example, in order to obtain the cybersecurity curricula for Austrian universities, the official database of studienwahl.at was used. This is the recommended database for prospective students in Austria maintained by the Austrian ministry of science and education \cite{FederalMinistryofScienceResearchandEconomy.2021}. The curricula for the Swiss and German universities were collected by checking each university's website and gathering the necessary information.

The relevant cybersecurity course information was identified in curricula of 13 Swiss, 12 Austrian, and 11 German universities. They were available in German or English. The data corpus comprises 842 different courses. The distribution of courses per country and language is shown in Table \ref{tab:curricula}. 

\begin{table}[h!]
\centering
\setlength\tabcolsep{1.5pt}
\begin{tabular}{|l|l|l|l|l|l|l|}
\hline
Country (language) & S (ger) & S (en) & A (ger) & A (en) & G (ger) & G (en) \\ \hline
Number of courses & 0 & 84 & 138 & 200 & 392 & 28 \\ \hline
\end{tabular}
\caption{Distribution of curricula per country and language}
\label{tab:curricula}
\vspace{-1.5\baselineskip}
\end{table}

\subsection{Manual Extraction of Skill Categories}
\label{subsec:manual_techniques}

Manual skill analysis relies on existing cyber-security taxonomies. We refer to a taxonomy as a hierarchical list of keyword sets describing specific cybersecurity skill categories. In this work, we chose the ACM CCS \cite{ACM} and the EU Cybersecurity Taxonomy \cite{JRC} because they were suitable for our approach described in Section~\ref{fuzzymatch}. These taxonomies are concise and have a clear hierarchy with 2 layers of skill categories. 

\subsubsection{ACM CCS}
The Association for Computing Machinery (ACM) proposed a Computing Classification System
(CCS) \cite{ACM} that includes the "Security and Privacy" category as a top area. The latest version of ACM CCS was updated in the year 2012. Its purpose is to classify publications submitted to ACM events. Table \ref{tab:ACM Taxon} shows an example of L1 and L2 layers  in ACM CCS.

\begin{table}[h!]
\centering
\setlength\tabcolsep{1.5pt}
\begin{tabular}{ |c|c|c| }
\hline
L1 & L2 \\
\hline
Cryptography &	Key management\\ & Public key cryptography\\ & Digital signatures\\ 
& Symmetric cryptography\\
& Block and stream ciphers\\ 
& Hash functions and message authentication codes\\
& Cryptanalysis and other attacks\\ 
& Information-theoretic techniques\\ 
& Mathematical foundations of cryptography \\
\hline
\end{tabular}
\caption{Exemplary L1 and L2 categories of ACM CCS}
\label{tab:ACM Taxon}
\vspace*{-1.5\baselineskip}
\end{table}

\subsubsection{EU Cybersecurity Taxonomy}
The main objective of the EU Cybersecurity Taxonomy (CST) developed by the EU Joint Research Center is to align cybersecurity terminologies, definitions, and domains used across different EU countries into a coherent and comprehensive taxonomy \cite{JRC}. This taxonomy contains three "dimensions": research domains, sectors and technologies, and use cases. Subcategories of research domains are most closely related to skills; therefore this dimension is selected as the base taxonomy for our analysis.

\subsubsection{Fuzzy Keyword Matching} \label{fuzzymatch}
The main task of our manual analysis is to search for specific skill categories from the respective taxonomies in job ads and curricula descriptions. Each skill category is represented by a set of keywords, e.g., \{"block", "and", "stream", "ciphers"\}. Since the taxonomies are developed in English, documents in a different language were automatically translated into English using the DeepL API. The search was performed using a Python library \textit{fuzzywuzzy} \cite{fuzzywuzzy}, which provides various algorithms for fuzzy string matching. Specifically, a token set matching approach is used in this work which is suitable for comparing strings that are vastly different in size. This method yields a similarity score in the range 0-100, where 100 corresponds to the exact match.

Finally, for each skill category, we compute the percentage of documents in the corpus for which the similarity score with this category's keyword set exceeds 90\%. The matching process is carried out for all skill categories in L2. The respective values for skill categories in L1 are aggregated over their subcategories. 

\subsection{Automatic Extraction of Skill Categories}
\label{subsec:automatic_techniques}

The goal of automatic extraction of skill categories is to identify sets of keywords that are characteristic of certain skills. This task is closely related to topic modeling in text analysis in which such sets of keywords are interpreted as topics addressed in a document. The latter problem belongs to the class of unsupervised machine learning and has been extensively addressed in the literature (cf. \cite{alghamdi2015survey} for a detailed review of related methods).

The best-known approach for topic modeling is the Latent Dirichlet Allocation (LDA) \cite{blei2003latent} in which each document is assumed to have a variety of topics represented by various keyword sets. Despite the success of LDA models for topic extraction from large datasets, they are language-specific and limited to a fixed vocabulary. Recent work in deep learning, especially the pre-trained Transformer models such as BERT \cite{devlin2018bert}, bring new opportunities for topic modeling. Pre-trained neural topic models replace keyword sets with learned contextual embeddings.  Transformer models can, furthermore, handle different languages via an unsupervised mapping between languages during pre-training.

To discover the topics in different corpora and languages, we use the contextual topic modeling approach of \citet{bianchi2021cross}, particularly suitable for cross-lingual settings. The main idea of this approach is to first create a topic model for a reference data corpus, and then to use zero-shot learning to infer similarly related topics in other corpora, potentially in different languages. In addition, we leverage a pre-trained NLP model for a different task, next sentence prediction, in order to create word embeddings in different languages and use them, instead of bag-of-words in a specific language, as input for topic modeling. We use the model
\texttt{distilbert-base-multilingual-cased} \cite{sanh2019distilbert}, trained on documents in 104 languages, including German and English, collected from Wikipedia. This model can be adapted to topic modeling via a standard transfer learning technique known as "fine-tuning".

 A crucial task of topic modeling that require special attention is determining the optimal number of topics. We repeatedly run the topic modeling step on our reference corpus, English job ads, for a various number of topics ranging from 5 to 50. The quality of topics is assessed using the Normalized Pointwise Mutual Information (NPMI) score which is closely correlated with human judgments of topic quality \cite{lau2014machine}.  According to this empirical evaluation, the most plausible number of topics in our English job ad corpus is 20, which is feasible as the number of key skills in cybersecurity. The resulting topics require interpretation since they comprise several dozen of keywords. The final topic annotation was carried out manually by three experts who assigned short descriptions based on the 30 most likely words reported for each topic.

\section{Identified Skill Gaps}
\label{sec:gaps}

In this section, we analyze the results of skill category extraction and identify skill gaps. As explained in Sections~\ref{subsec:manual_techniques} and \ref{subsec:automatic_techniques}, both analysis methods return the document frequency of skill categories.  
 If a substantial difference in the document frequency is observed for some skill category this is reported as a skill gap.

\subsection{Manual Analysis Results}
\label{subsec:manual}

\begin{figure}[t]
    \centering
    \includegraphics[width=\columnwidth]{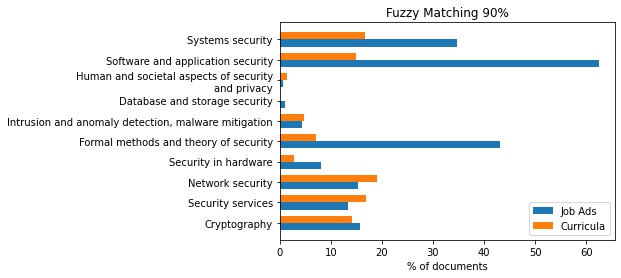}
    \caption{L1 skill gap analysis (ACM CCS)}
    \label{fig:ACM-fuzzy-90}
    \vspace{-0.5\baselineskip}
\end{figure}

\begin{figure}[t]
    \centering
    \includegraphics[width=\columnwidth]{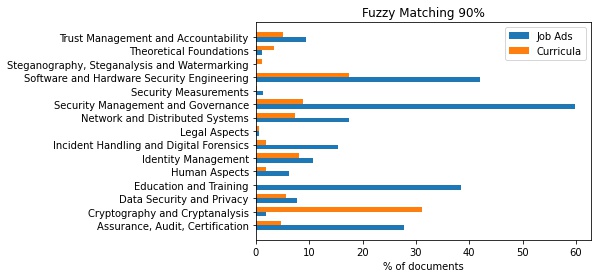}
    \caption{L1 skill gap analysis (EU CST)}
    \label{fig:JRC-fuzzy-90}
\end{figure}

We first present the manual skill gap analysis results at the L1 level for the two taxonomies, shown in Figures \ref{fig:ACM-fuzzy-90} and \ref{fig:JRC-fuzzy-90}, respectively. These figures display the percentage of documents ``supporting'' the respective skill categories in job ads and academic curricula.

Clear skill gaps can be observed for several L1 skill categories. Substantial discrepancies in ``support'' exist for the skill categories "Systems Security", "Software and Application Security", "Formal Methods and Theory of Security" and "Security Hardware" in the ACM CCS and for the skill categories "Software and Hardware Security Engineering", "Security Management and Governance", "Network and Distributed Systems", "Incident Handling and Digital Forensics", "Assurance, Audit and Certification" of JRC taxonomy.

A deeper understanding of identified skill gaps can be obtained by analyzing results at L2. The results cannot be presented here for all L2 categories\footnote{Comprehensive results can be found at \url{https://tinyurl.com/midsise}.}, yet some examples of such analysis presented below are very instructive.

\begin{figure}[t]
    \centering
    \includegraphics[width=\columnwidth]{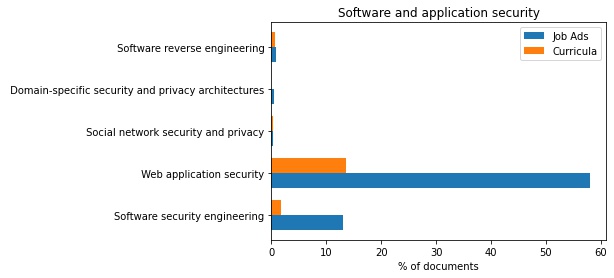}
    \caption{Software and application security (ACM CCS)}
    \label{fig:ACM-Software-90}
\end{figure}

Figure \ref{fig:ACM-Software-90} shows a more detailed L2 skill gap analysis for the L1 category ``Software and Application Security'' in ACM CCS for which a large skill gap is observed in Figure~\ref{fig:ACM-fuzzy-90}.
Here we can see that the largest skill gaps exist for the subcategories ``Web Application Security'' and ``Software Security Engineering''. Such topics are in high demand in today's IT industry, whereas academic offerings seem to insufficiently address these skills. 

\begin{figure}[tb]
    \centering
    \includegraphics[width=\columnwidth]{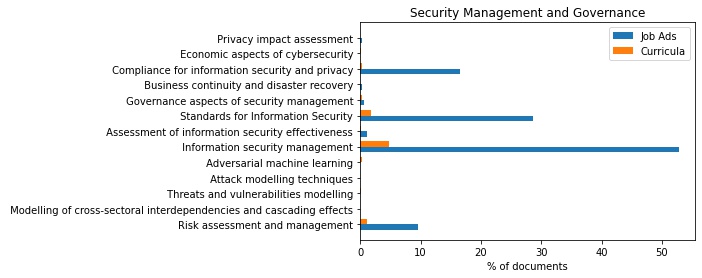}
    \caption{Security management and governance (EU CST)}
    \label{fig:JRC-Management-90}
    \vspace{-0.5\baselineskip}  
\end{figure}

Another example illustrates the skill gap in the L1 category  ``Security Management and Governance'' of the EU CST, shown in Figure \ref{fig:JRC-Management-90}. One can see four prominent skill gaps related to information security management, risk assessment, compliance and standards. All these skills are under-represented in the academic curricula. 

\begin{figure}[tb]
    \centering
    \includegraphics[width=\columnwidth]{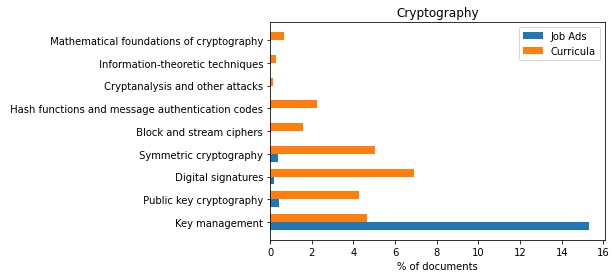}
    \caption{L2 analysis for ``Cryptography'' (ACM CCS)}
    \label{fig:ACM-Crypto-90}
        \vspace{-0.5\baselineskip} 
\end{figure}

\begin{figure}[tb]
    \centering
    \includegraphics[width=\columnwidth]{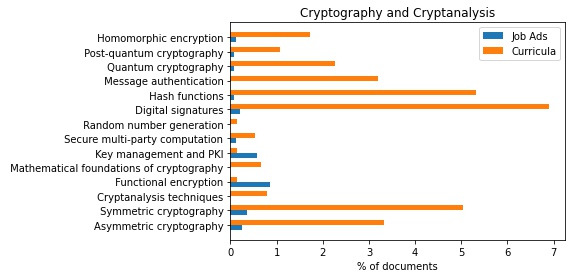}
    \caption{L2 analysis for ``Cryptography and Cryptanalysis'' (EU CST)}
    \label{fig:JRC-Crypto-90}    \vspace{-0.5\baselineskip} 
\end{figure}

In general, one cannot expect a full coherence between the results based on the two taxonomies due to their differences in the specific composition of L1 and L2 categories. Even seemingly contradictory results may be observed which, however, make sense after a closer investigation. For example, at L1, no skill gap is observed in the category ``Cryptography'' in ACM CCS but a negative skill gap is observed in EU CST. The L2 analysis shown in Figures \ref{fig:ACM-Crypto-90} and \ref{fig:JRC-Crypto-90} reveals, however, that a negative skill gap \emph{does} exist in almost all L2 categories in both taxonomies but is compensated for the ACM CCS by the skill gap in the category ``Key Management'' which is much more general an hence more prominent among job ads.

\subsection{Automatic Analysis}
\label{subsec:automatic}

\begin{figure}[tbp]
    \centering
    \includegraphics[width=0.9\columnwidth]{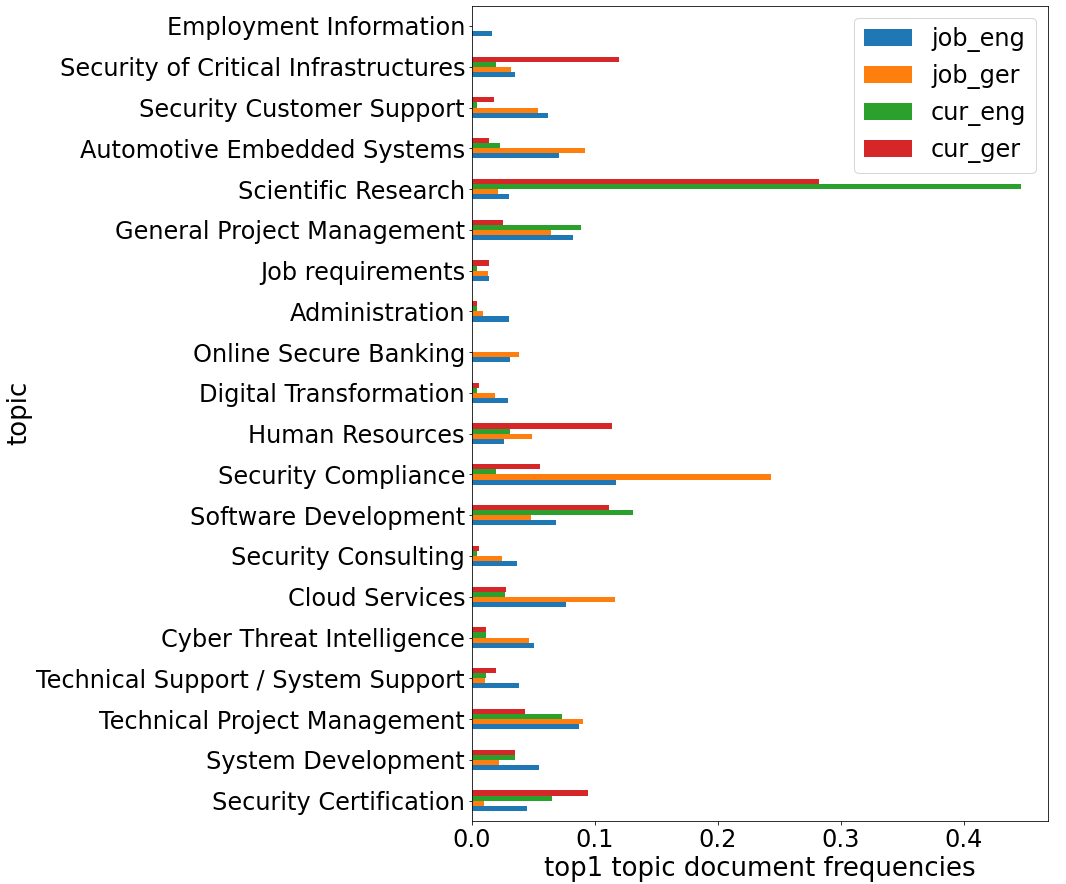}
    \caption{Automatic skill gap analysis.}
    \label{fig:CTM_results}
\end{figure}

The results of the automatic skill gap analysis are presented in Figure~\ref{fig:CTM_results}. It shows document frequencies for 20 skill categories identified as topics by the text mining technique presented in Section~\ref{subsec:automatic}. For each topic, document frequencies are presented for job ads and curricula, in English and in German. 

The obtained results offer some intriguing insights about the nature of skill gaps as well as the analysis technique itself. The first interesting observation is that the identified topics not only represent the skills but also other information typical for job ads, e.g., employment information, administration, job requirements, and HR-related information.  Another interesting fact is that some topics represent IT branches rather than skills, e.g., automotive embedded systems and online banking. This effect is quite natural and indicative of specific branches where the demand for cybersecurity skills is particularly strong. 

The specific skill gaps identified by our analysis are very plausible. The strongest discrepancy between job ads and curricula in Figure~\ref{fig:CTM_results} can be observed for security compliance, followed by cloud services, automotive embedded systems, online banking, and threat intelligence. Some other observed discrepancies, e.g., in security customer support, security consulting, and the ``inverse'' skill gap in scientific research are also quite obvious. It noteworthy that the skill gaps identified by the automatic analysis are largely consistent with the manual analysis, except that cloud security was only discovered by the automatic analysis. To our surprise, cloud security is mentioned neither in the ACM CCS nor in the EU Cybersecurity Taxonomy, which corroborates our observation that skill gaps must be periodically re-assessed.

\section{Discussion}
\label{sec:discussion}

Both automatic and manual extraction of skill categories can accurately infer skills required in industry and produced by the academia. Despite the relatively small size of our data corpora as well as its regional focus on German-speaking countries, the identified skill gaps are plausible and enable concrete curricula development recommendations, as elucidated in the following section. 

The multilingual nature of job ads and curricula descriptions can be addressed in two different ways. For the manual analysis built on existing taxonomies, automatic translation must be used in order to convert the data into the target language of the taxonomy. This feature is indispensable for smaller regional markets with no skill taxonomies available beforehand. Automatic analysis building on multilingual NLP techniques, in contrast, exhibits a strong potential both for regional markets and in the global international context. 

The developed methodology is clearly applicable to other disciplines in computer science beyond cybersecurity. The broad availability of job ad and curricula data  makes the adaptation of out methods to other disciplines rather straightforward. Our methods are also suitable for a periodic re-assessment of existing skill gaps, especially as a way to measure the impact of educations development. Our tools are available as an open source code. \footnote{\url{https://tinyurl.com/EMIDSISE}}

\section{Recommendations for curricula design}

\begin{figure}[tbhp]
    \centering
    \includegraphics[width=1\columnwidth]{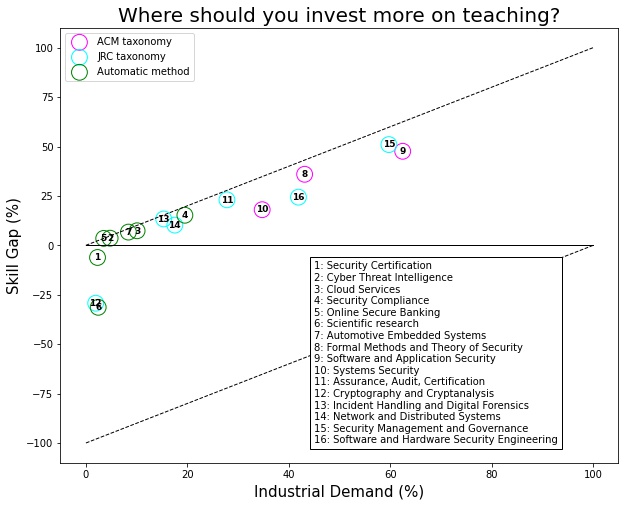}
    \caption{Summary and prioritization of identified skill gaps}
    \label{fig:Global_gartner}
\end{figure}

In order to support the design of future curricula in cybersecurity that are capable of meeting industrial needs, we have distilled all our results into a new graph representation shown in Figure~\ref{fig:Global_gartner}. Its horizontal axis represents the industrial demand (measured as the percentage job ads), and its vertical axis represents the strength of a skill gap (measured as the difference in the percentages of job ads and curricula ``supporting'' the respective skill category). Skill categories that are close to the upper diagonal line exhibit strong skill gaps; they are further prioritized (from left to right) by the absolute strength of industrial demand. The negative skill gaps, indicative of the oversupply of academic programs, can be seen as entries in the negative part of the graph in Figure~\ref{fig:Global_gartner}. The results of three different analysis techniques presented in the previous sections are shown in different colors. 

The following specific recommendations can be articulated derived from  Figure~\ref{fig:Global_gartner}. More courses covering security management and governance as well as software and application security should be added to academic curricula. The discrepancy between the industrial demand and the academic supply is very strong in these two areas. Also formal methods and theory of security (which covers security requirements engineering) as well as software and hardware security engineering have a high industrial demand and a substantial skill gap, and hence should be better addressed in the future curricula. In contrast, scientific research as well as cryptography and cryptanalysis seem to be over-represented in curricula, and there is, in general, a low industrial demand on skills in these areas. The recommendations derived from Figure~\ref{fig:Global_gartner} are in line with the results from both the manual and the automatic analysis discussed before. Some emerging skill categories such as security compliance, cloud services, cyber threat intelligence and automotive embedded systems, in which moderate skill gaps were identified by our automatic analysis, should also be considered as potential ``skill bottlenecks'' in the near future.

\section{Conclusions}
\label{sec:conclusions}

The methods presented in this paper as well as the results of the skill gap analysis obtained with the help of these methods constitute an important step towards the understanding of skill gaps in specific branches of the IT industry. Perhaps the most important lesson we learned from this study is that skill gaps are highly dynamic. Despite a substantial effort invested worldwide into the systematization of cybersecurity knowledge, the skill landscape in the IT industry closely follows fast-evolving technology trends.  Fixed skill taxonomies seem to be insufficient for capturing such development and hence, the automatic analysis techniques presented in our paper are certainly a promising instrument, especially for a periodic re-assessment of the skill landscape. 

Our findings certainly support the claim that cybersecurity problems cannot be solved by technical solutions alone. Hence, the technical part of cybersecurity education should be complemented by stronger offerings on the managerial side where a strong skill gap can be observed in our current results. Numerous gaps also exist in the technical disciplines related to cybersecurity, especially in cloud security, threat intelligence, incident response, and forensics.  

\section*{Acknowledgements}

The funding for this work was provided by the EU Erasmus+ Programme under the Grant Agreement Number 2018-1-LI01-KA203-000114 (project MIDSISE). The authors are grateful to Ulrike Meyer (RWTH Aachen) and Fabio Di Franco (ENISA) for valuable feedback on the project results. 

\appendix

\bibliographystyle{ACM-Reference-Format}
\bibliography{references.bib}

\end{document}